\documentclass[epj]{svjour}
\usepackage{graphics}
\usepackage{url}
\usepackage{color}
\usepackage{mathrsfs}
\usepackage{amssymb}
\usepackage{bm}
\usepackage[numbers]{natbib}
\usepackage{rotating}
\usepackage[normalem]{ulem}
\usepackage{epstopdf}
\usepackage{lipsum}
\usepackage{mathtools}
\usepackage{cuted}
\usepackage{verbatim}
\usepackage[colorlinks,citecolor=blue,linkcolor=blue,anchorcolor=blue,filecolor=blue,
urlcolor=blue]{hyperref}
\usepackage{enumitem}

\begin{document}

\title{Thermodynamical phases in a PNJL model at zero temperature}

\author{O. A. Mattos\inst{1}, T. Frederico\inst{1}, and O. Louren\c{c}o\inst{1}}  

\institute{Departamento de F\'isica, Instituto Tecnol\'ogico de Aeron\'autica, DCTA, 12228-900, 
S\~ao Jos\'e dos Campos, SP, Brazil}

\date{\today}

\abstract{
The confinement/deconfinement transition described the Polyakov-Nambu-Jona-Lasinio (PNJL) model is 
extended to be operative at zero temperature regime. In this study, the scalar and vector channel 
interaction strengths of the original PNJL model are modified by introducing a dependence  on the 
traced Polyakov loop. In such a way the effective interactions depend on the quark phase and in 
turn provides a backreaction of the quarks to the gluonic sector, also at zero temperature. On 
general grounds from quantum chromodynamics this is an expected feature. The thermodynamics of the 
extended model (PNJL0) is studied in detail. It presents along with a suitable choice of the 
Polyakov potential, a first order confined/deconfined quark phase transition even at $T=0$. We also 
show that the vector channel plays an important role in order to allow $\Phi\ne0$ solutions for the 
PNJL0 model. Furthermore, the sensitivity of the combined quarkyonic and deconfinement phases 
to the vector interaction strength and the proposed parametrization of the Polyakov-loop potential 
at $T=0$ allowed to set a window for the bulk values of the relevant parameters. 
}


\maketitle

\section{Introduction} 

The physics of strongly interacting matter is described by Quantum Chromodynamics 
(QCD)~\cite{Weinberg, Fritzsch, Huang} with quarks and gluons as the fundamental degrees of freedom, 
while the colorless
hadrons are  emergent phenomena originated from the complex nonperturbative physics of this theory. 
Despite being well established, nonperturbative aspects of QCD 
have nontrivial treatment is far from being understood, which is critical for the description of
the equation of state  of strongly interacting matter at finite densities and low temperatures. 

The challenge is to compute hadronic observables when the strength of the QCD  running coupling presents 
the well known strong infrared enhancement to embody spontaneous chiral symmetry breaking and 
at the same time  quark and gluon confinement.
These properties are opposite  to what is know in Quantum Electrodynamics (QED),
where the use of perturbative methods are successful and the
fundamental degrees of freedom appears as asymptotic states. That  is not the case for QCD 
where quarks and gluons are confined. 

Ab-initio nonperturbative approaches are called for to solve QCD, as Lattice Quantum 
Chromodynamics (LQCD) \cite{Kogut1,Kogut2,Rothe}. This well known method 
starts with the QCD action, $S_{\mbox{\tiny QCD}}$, and evaluate the generating 
functional $Z=\int \mathcal{D}A\mathcal{D}\psi\mathcal{D}\bar{\psi} e^{iS_{\mbox{\tiny  QCD}}}$ by
numerical simulations and finally accessing the  matrix elements of the relevant operators between 
hadronic states.
Such a procedure is performed through both, space-time discretization and Wick rotation, and changes 
$Z$ to $Z'=\int \mathcal{D}A\mathcal{D}\psi\mathcal{D}\bar{\psi} e^{-S_{\mbox{\tiny  QCD}}}$, i.e., 
putting it in correspondence to a Statistical Mechanics formulation. 
The numerical calculations are performed from Monte Carlo simulations. 
Despite powerful, LQCD faces some intrinsic problems such as the need to 
extrapolation of the outcomes for lattice spacing approaching to zero, the need of powerful 
dedicated computational facilities, and the ``fermion sign problem''~\cite{Muroya}.

Another nonperturbative continuum method to treat QCD is through Dyson-Schwinger equations 
(DSE)~\cite{Roberts,Alkofer}. They are derived from the generating functional $Z$ and 
allows to obtain the  equations of motion of the n-point functions, which 
 are also known as the Euler-Lagrange equations for the QCD Green's function. 
 Both LQCD and the standard DSE methods have the issue of being formulated in the Euclidean 
space. However, the access of all observables obtained from  QCD, requires its representation 
in the Minkowski space, that calls for a careful and not yet fully known analytical extension or
other particular methods to obtain, for example, the light-front wave function of the hadronic state. 
A possible alternative to circumvent this problem is the use of the Nakanishi integral representation, 
built directly in the Minkowski space, and solve with that the DSE or even the Bethe-Salpeter 
equation~\cite{Paula,Frederico,Pimentel}. Sum rules~\cite{Shifman, REINDERS}, and the connection 
between gauge and string theories also aim to treat  QCD in the infrared region. This last method was 
introduced by Gerard 't Hooft~\cite{HOOFT} by suggesting the analytical calculation of amplitudes 
by using string theories. 
A clear mapping 
between the two theories was proposed by Juan Maldacena~\cite{MALDACENA} with the famous Anti-de 
Sitter/Conformal Field Theory (AdS/CFT) conjecture.

Other practical possibilities to incorporate the infrared physics is the use of effective quark 
models, built with the aim of reproducing  most of the QCD phenomenology, as for instance,  their 
symmetries and also dynamical chiral symmetry breaking. By exploiting such approaches, many models 
were developed. The Nambu-Jona-Lasinio (NJL) 
model~\cite{Nambu1,Nambu2,buballa,Vogl,Klevansky,Hatsuda,ric1,ric2,ric3} is an example. Its improved version, 
namely, the Polyakov-Nambu-Jona-Lasinio (PNJL) 
model~\cite{FUKUSHIMA,fukushima3,fuku1,fuku2,weise1,weise2,weise4,weise6,ratti,costa,scoccola,
nosso1,nosso2,nosso3,nosso4} includes effects of confinement / deconfinement phase transition, 
included in the theory through an effective field of gluonic origin, namely, the Polyakov loop 
($\Phi$ and $\Phi^*$)~\cite{POLYAKOV, SUSSKIND, SVETITSKY1, SVETITSKY2}. Even that the PNJL model 
shows a clear improvement in comparison with the original NJL model, with respect to QCD phase 
structure, the Polyakov loop decouples from the baryonic fields at the zero temperature regime. 
This missing interaction is evident by analyzing, for instance, the pressure ($P$) and the energy 
density ($\mathcal{E}$) of PNJL models at low temperatures. The Polyakov potential, 
$\mathcal{U}(\Phi,\Phi^*,T)$, vanishes for $T=0$ in the most parametrizations of this quantity. 
Therefore, the equations of state of the NJL model are recovered at $T=0$ and, consequently, the 
physics of confinement  implemented in $ T\neq 0$ in PNJL models is simply lost. 

In this work we  explore the approach started with the initial study of 
Ref.~\cite{mattos} and present the thermodynamics of a PNJL model at zero temperature, 
named here as PNJL0 model. In the following, Sec.~\ref{secpnjl}, we briefly 
review the basics of the original PNJL model. Its version at $T=0$, along with its main results, 
are discussed in Sec.~\ref{secpnjl0}. Finally, we present the summary and concluding remarks of our 
study in Sec.~\ref{secsummary}.

\section{Polyakov-Nambu-Jona-Lasinio model}
\label{secpnjl}

The PNJL model was firstly proposed in Ref.~\cite{FUKUSHIMA} as a generalization of the original NJL 
model due to the inclusion of confinement effects. From this point of view, the PNJL model becomes 
an effective model describing the QCD theory more realistically in comparison with the previous NJL 
version. Basically, the gluon dynamics is implemented in the NJL model by replacing the derivative 
$\partial^\mu$ by $D^\mu\equiv\partial^\mu+iA^\mu $ where $A^\mu=\delta^\mu_0A_0$ 
and $A_0=gA_\alpha^0\lambda_\alpha / 2$ ($g$ is the gauge coupling and $\lambda_\alpha
$ are the Gell-Mann matrices). The Lagrangian density of the SU(2) PNJL model is then written as
\begin{eqnarray}
\mathcal{L}_{\mbox{\tiny PNJL}} &=& \bar{\psi}(i\gamma_\mu D^\mu - m)\psi 
+G_s\left[(\bar{\psi}\psi)^2-(\bar{\psi}\gamma_5 \tau\psi)^2 \right]  \nonumber\\
&-&G_V(\bar{\psi}\gamma_\mu\psi)^2- \mathcal{U}(\Phi,\Phi ^*,T).\qquad
\label{dlpnjl}
\end{eqnarray}
with $m$ being the current quark mass (in our case $m=m_u=m_d$). In this formulation, we also add a 
vector channel, regulated by the coupling constant $G_V$ . 

Other clear difference between PNJL and NJL models is the inclusion of the Polyakov potential 
$\mathcal{U}(\Phi,\Phi ^*,T)$ that depends on the traced Polyakov loop and its conjugate, $\Phi$ 
and $\Phi^*$, respectively. 
$\Phi$ is defined in terms of $A_4=iA_0\equiv T\phi$ as
\begin{align}
\Phi&=\frac{1}{3}\rm{Tr}\left[\,\,\rm{exp}\left(i\int_0^{1/T}d\tau\,A_4\right)\right]
\nonumber \\
&=\frac{1}{3}\rm{Tr}\left[\rm{exp}(i\phi)\right]
=\frac{1}{3}\rm{Tr}\left\lbrace\rm{exp}[i(\phi_3\lambda_3+\phi_8\lambda_8)]
\right\rbrace \nonumber \\
&=
\frac{1}{3}\left[\rm{e}^{i(\phi_3+\phi_8/\sqrt{3})}+\rm{e}^{i(-\phi_3+\phi_8/\sqrt{3})}
+\rm{e}^{-2i\phi_8/\sqrt{3}}\right],
\label{traced}
\end{align}
in a gauge (Polyakov gauge) in which the gluon field is written in terms of the diagonal
Gell-Mann matrices as $\phi=\phi_3\lambda_3+\phi_8\lambda_8$, with $\phi_3,\phi_8 \in 
\mathbb{R}$ (the definitions $\phi_3=A_4^3/T$ and $\phi_8=A_4^8/T$ were taken into account). Here 
we also use the mean-field approximation described in Refs.~\cite{weise4,weise6,nosso3} that 
considers the mean-field configuration in which $\phi_8=0$ in Eq.~(\ref{traced}). In this case, 
$\Phi=\Phi^*=[2\cos(\phi_3)+1]/3$ even for nonvanishing quark chemical potentials ($\mu$).

The thermodynamics of this model is obtained through the calculation of its grand canonical 
potential density 
namely, $\Omega_{\mbox{\tiny PNJL}} = -T\mbox{ln} (Z_{\mbox{\tiny PNJL}})/V$, 
with $Z_{\mbox{\tiny PNJL}}$ being the partition function of the model. The final 
expression is given by~\cite{FUKUSHIMA,weise1,weise2,rossner1,rossner2}
\begin{eqnarray}
\Omega_{\mbox{\tiny PNJL}} &=& \mathcal{U}(\Phi,T) +
G_s\rho_s^2 - G_V\rho^2- \frac{\gamma}{2\pi^2}\int_0^{\Lambda}E\,k^2dk \nonumber \\
&-& \frac{\gamma T}{2\pi^2N_c}\int_0^{\infty}\mbox{ln}\Big[1+3\Phi e^{-(E - \mu)/T} \nonumber\\
&+& 3\Phi e^{-2(E - \mu)/T} + e^{-3(E - \mu)/T} \Big]k^2dk \nonumber\\
&-& \frac{\gamma T}{2\pi^2N_c}\int_0^{\infty}\mbox{ln}\Big[1+3\Phi e^{-(E + \mu)/T} \nonumber\\
&+& 3\Phi e^{-2(E + \mu)/T} + e^{-3(E + \mu)/T} \Big]k^2dk,
\label{omegapnjl}
\end{eqnarray}
with $E=(k^2+{M}^2)^{1/2}$, $\rho_s=\left<\bar{\psi}\psi\right>=\left<\bar{u}u\right> + 
\left<\bar{d}d\right>=2\left<\bar{u}u\right>\,$ and the degeneracy factor given by 
$\gamma=N_s\times N_f\times N_c=12\,$, due to the spin, flavor and color numbers, respectively 
($N_s=N_f=2$ and $N_c=3$). The quantity $\Lambda $ defines the cutoff of the divergent integral. 
As in the NJL model, the constituent quark mass $M$ is given in terms of the quark condensate 
$\rho_s$ as
\begin{eqnarray}
M=m-2G_s\rho_s,
\label{massapnjl}
\end{eqnarray}
with $\rho_s$, obtained by the condition $\partial\Omega_{\mbox{\tiny PNJL}}/\partial\rho_s=0$, 
given by
\begin{eqnarray}
\rho_s &=& \frac{\gamma}{2\pi^2}\int_0^{\infty}\frac{M}{E(M)}k^2dk 
\left[ f(k,T,\Phi) + \bar{f}(k,T,\Phi) \right] \nonumber\\
&-&\frac{\gamma}{2\pi^2}\int_0^{\Lambda}\frac{M}{E(M)}k^2dk.
\label{rhospnjl}
\end{eqnarray}
The functions $f(k,T,\Phi)$ and $\bar{f}(k,T,\Phi)$, given as follows,
\begin{align}
&f(k,T,\Phi)=\nonumber\\
&\frac{\Phi e^{2(E-\mu)/T} + 2\Phi e^{(E-\mu)/T}+ 1}
{3\Phi e^{2(E-\mu)/T} + 3\Phi e^{(E-\mu)/T} + e^{3(E-\mu)/T} + 1}
\label{fdmp} 
\end{align}
and
\begin{align}
&\bar{f}(k,T,\Phi) =\nonumber\\
&\frac{ \Phi e^{2(E+\mu)/T}+2\Phi e^{(E+\mu)/T}+1}
{ 3\Phi e^{2(E+\mu)/T} + 3\Phi e^{(E+\mu)/T} + e^{3(E+\mu)/T} + 1},
\label{fdmap}
\end{align}
are the generalized Fermi-Dirac distributions, also used to obtain the quark density through
\begin{eqnarray}
\rho &=& -\frac{\partial \Omega_{\mbox{\tiny PNJL}}}{\partial\mu} 
\nonumber\\
&=& \frac{\gamma}{2\pi^2}\int_0^{\infty}k^2dk[f(k,T,\Phi) - \bar{f}(k,T,\Phi)].\quad \,
\label{rhopnjl}
\end{eqnarray}

Note the similarity between the grand canonical potentials of the PNJL and NJL models, where in the 
former there is the replacement of the usual Fermi-Dirac functions of quarks and antiquarks by the 
generalized functions given in Eqs~(\ref{fdmp}) and~(\ref{fdmap}). Furthermore in the PNJL model, 
there is also the inclusion of an effective gluon  potential represented by $\mathcal{U}(\Phi,T)$ 
in the grand canonical potential density. 

The effective scalar field $\Phi$ is found through the solution of $\partial\Omega_{\mbox{\tiny 
PNJL}}/\partial\Phi=0$. This quantity is determined simultaneously to $M$, that is found from 
Eqs.~(\ref{massapnjl}) and~(\ref{rhospnjl}), or equivalently, through the condition 
$\partial\Omega_{\mbox{\tiny PNJL}}/\partial\rho_s=0$. Pressure and energy density are obtained from 
Eq.~(\ref{omegapnjl}) as
\begin{align}
&P_{\mbox{\tiny PNJL}}(\rho,T) = -\Omega_{\mbox{\tiny PNJL}} =
-\mathcal{U}(\Phi,T) + G_V\rho^2 - G_s\rho_s^2 
\nonumber \\
&+ \frac{\gamma}{2\pi^2}\int_0^{\Lambda}(k^2+M^2)^{1/2}\,k^2dk + \mathcal{E}_{\rm vac}
\nonumber \\
&+ \frac{\gamma}{6\pi^2}\int_0^{\infty}\frac{k^4}{(k^2+M^2)^{1/2}}dk[f(k,T,\Phi)+\bar{f}(k,T,\Phi)]
\label{ppnjl}
\end{align}
and
\begin{align}
&\mathcal{E}_{\mbox{\tiny PNJL}}(\rho,T) = -T^2\frac{\partial (\Omega_{\mbox{\tiny 
PNJL}}/T)}{\partial T} 
+ \mu\rho \nonumber\\
&= \mathcal{U}(\Phi,T) - T\frac{\partial\mathcal{U}}{\partial T} + G_V\rho^2 + G_s\rho_s^2 
\nonumber \\
&-\frac{\gamma}{2\pi^2}\int_0^{\Lambda}(k^2+M^2)^{1/2}\,k^2dk - \mathcal{E}_{\rm vac}
\nonumber \\
&+ \frac{\gamma}{2\pi^2}\int_0^{\infty}(k^2+M^2)^{1/2}\,k^2dk [f(k,T,\Phi) + \bar{f}(k,T,\Phi)],
\label{depnjl}
\end{align}
respectively, with the vacuum quantity $\mathcal{E}_{\rm vac}$ included in the equations in order to 
ensure $P=\mathcal{E}=0$ at vanishing temperature and quark density. Finally, the entropy density 
can be obtained from $\mathcal{S}_{\mbox{\tiny PNJL}} = -\partial\Omega_{\mbox{\tiny PNJL}}/\partial 
T$, or from $\mathcal{S}_{\mbox{\tiny PNJL}}=(P_{\mbox{\tiny PNJL}} + \mathcal{E}_{\mbox{\tiny 
PNJL}} - \mu\rho)/T\,$. The thermodynamics of the PNJL model is quantitatively defined once the 
potential $\mathcal{U}(\Phi,T)$, and the constants $G_s$, $\Lambda$ and $m$ are chosen. These last 
quantities are the same as the ones obtained in the quarks sector (NJL model), with $ G_V $ being a 
free parameter.

\section{PNJL model at zero temperature (PNJL0)}
\label{secpnjl0}

\subsection{Construction of the model}

It is worth notice that the limit of vanishing temperature in Eqs.~(\ref{ppnjl}) 
and~(\ref{depnjl}) leads to
\begin{align}
P_{\mbox{\tiny PNJL}}&(\rho,0) = G_V\rho^2-G_s\rho^2_{s} +\frac{\gamma}{2\pi^2}\int^{\Lambda}_{0} 
dk k^2(k^2+M^2)^{1/2} 
\nonumber\\
&+ \frac{\gamma}{6\pi^2}\int_{0}^{k_F}dk \frac{k^4}{(k^2+M^2)^{1/2}} + \mathcal{E}_{\rm 
vac} =-\Omega_{\mbox{\tiny PNJL}}(\rho,0)
\label{pressnjl} 
\end{align}
and
\begin{eqnarray}
\mathcal{E}_{\mbox{\tiny PNJL}}(\rho,0) &=& G_V\rho^2 + G_s\rho^2_{s} - \mathcal{E}_{\rm vac}
\nonumber\\
&-& \frac{\gamma}{2\pi^2}\int_{k_F}^{\Lambda}dk k^2(k^2+M^2)^{1/2},
\label{denjl}
\end{eqnarray}
with $\mu = 2G_V\rho + (k_F^2+M^2)^{1/2}$, i.e., at $T=0$ one has $P_{\mbox{\tiny 
PNJL}}(\rho)=P_{\mbox{\tiny NJL}}(\rho)$ and $\mathcal{E}_{\mbox{\tiny 
PNJL}}(\rho)=\mathcal{E}_{\mbox{\tiny NJL}}(\rho)$, with $P_{\mbox{\tiny NJL}}(\rho)$ and 
$\mathcal{E}_{\mbox{\tiny NJL}}(\rho)$ being the pressure and energy density, respectively, of the 
original NJL model at zero temperature~\cite{Nambu1,Nambu2,buballa,Vogl,Klevansky,Hatsuda}. Therefore, the 
confinement physics from the Polyakov potential is lost at $T=0$. Such  problem is due to two reasons. 
The first one is that the generalized Fermi-Dirac distributions, Eqs.~(\ref{fdmp})-(\ref{fdmap}),  
becomes the traditional step function $\theta(k_F-k)$ at $T=0$ ($k_F$ is the quark Fermi momentum). 
The second reason is  that the gluonic contribution of the PNJL model, described by the Polyakov 
potential, $\mathcal{U}(\Phi,T)|_{T=0}$ vanishes in the most known versions. 

Three most known forms of the Polyakov potential, namely, RTW05~\cite{weise1}, 
RRW06~\cite{weise2,weise4} and FUKU08~\cite{fukushima3}, are given respectively by $(\Phi=\Phi^*)$,
\begin{align}
\frac{\mathcal{U}_{\mbox{\tiny RTW05}}}{T^4} &= -\frac{b_2(T)}{2}\Phi^2
- \frac{b_3}{3}\Phi^3 + \frac{b_4}{4}\Phi^4,  
\label{rtw05} \\
\frac{\mathcal{U}_{\mbox{\tiny RRW06}}}{T^4} &= -\frac{b_2(T)}{2}\Phi^2
+ b_4(T)\mbox{ln}(1 - 6\Phi^2 + 8\Phi^3 - 3\Phi^4), 
\label{rrw06} \\
\frac{\mathcal{U}_{\mbox{\tiny FUKU08}}}{b\,T} &= -54e^{-a/T}\Phi^2 
+ \mbox{ln}(1 - 6\Phi^2 + 
8\Phi^3 - 3\Phi^4), 
\label{fuku08}
\end{align}
with
\begin{eqnarray}
b_2(T) = a_0 + a_1\left(\frac{T_0}{T}\right) + a_2\left(\frac{T_0}{T}\right)^2 
+ a_3\left(\frac{T_0}{T}\right)^3,
\label{b2t}
\end{eqnarray}
and
\begin{eqnarray}
{b_4(T) = b_4\left(\frac{T_0}{T}\right)^3.}
\end{eqnarray}
These potentials are constructed to reproduce data from lattice calculations of the pure gauge 
sector concerning the temperature dependence of $\Phi$ and its first order phase transition, 
characterized by the jump of $\Phi$ from zero to a finite value at $T_0=270$~MeV. In 
Eqs.~(\ref{rtw05})-(\ref{fuku08}), $a$, $b$, $a_0$, $a_1$, $a_2$, $a_3$, $b_3$ and $b_4$ are 
dimensionless free parameters. Notice that for all potentials one has $\mathcal{U}=0$ at $T=0$. This 
phenomenology leads the thermodynamics of the PNJL model to be reduced to the NJL one at zero 
temperature. One exception to this result is the DS10~\cite{Schramm1,Schramm2} potential given by 
\begin{eqnarray}
\mathcal{U}_{\mbox{\tiny DS10}} &=& (a_0T^4 + a_1\mu^4 + a_2T^2\mu^2)\Phi^2 + \mathcal{U}_0(\Phi)
\label{ds10}
\end{eqnarray}
with
\begin{eqnarray}
\mathcal{U}_0(\Phi)\equiv a_3T_0^4\mbox{ln}(1-6\Phi^2+8\Phi^3-3\Phi^4).
\label{u0}
\end{eqnarray}

Our aim is to avoid the lack of confinement physics in the PNJL model at $T=0$ by taking into account 
the effects of the traced Polyakov loop $\Phi$ in both the Polyakov potential and the effective 
interaction between the quarks,
as performed in a previous initial investigation~\cite{mattos} where preliminary results were presented. 
The idea here is to  introduce the traced Polyakov loop in the NJL equations of state by imposing 
vanishing scalar and vector couplings as the quarks become deconfined, situation predicted to occur 
at $\Phi\rightarrow 1$. This phenomenology can be achieved by making the scalar and vector coupling 
strengths dependent on $\Phi$ as follows,
\begin{eqnarray}
G_s \longrightarrow  G_s(\Phi) &=& G_s( 1-\Phi^2 ),
\label{gsphi} 
\end{eqnarray}
and
\begin{eqnarray}
G_V \longrightarrow  G_V(\Phi) &=& G_V( 1-\Phi^2 ).
\label{gvphi}
\end{eqnarray}
In fact, such changes can be seen as a simpler version of the Entanglement PNJL 
(EPNJL)~\cite{epnjl}. The implementation of Eqs.~(\ref{gsphi})-(\ref{gvphi}) in the PNJL/NJL model 
at $T=0$ still demands the determination of the possible $\Phi$ values, obtained through 
$\partial\Omega_{\mbox{\tiny PNJL}}/\partial\Phi=0$. However, the replacement of $G_s$ and $G_V$ by 
their $\Phi$ dependent versions is not enough to ensure $\Phi\ne 0$ solutions, i.e., the NJL model 
is recovered once more. In order to avoid this problem, besides the modifications proposed in 
Eqs.~(\ref{gsphi})-(\ref{gvphi}) we also add to $\Omega_{\mbox{\tiny PNJL}}(\rho,0)$ the 
term $\mathcal{U}_0(\Phi)$ given by Eq.~(\ref{u0}), inspired by the $\mathcal{U}_{\mbox{\tiny 
DS10}}$ Polyakov potential, with  $T_0 = 190$~MeV (value very often used in the Polyakov 
potentials of the PNJL models~\cite{weise1}). 

As we will discuss later on, the effect of 
$\mathcal{U}_0(\Phi)$ is to ensure $\Phi\ne 0$ solutions and also limit the traced Polyakov loop in 
the range of $0\leqslant\Phi\leqslant 1$. This term was used in Refs.~\cite{Schramm1,Schramm2} also to 
generate $\Phi\ne 0$, however, for a much more sophisticated model, that takes into account hadrons 
and quarks degrees of freedom at the same Lagrangian density. 

The quark couplings given by Eqs.~(\ref{gsphi})-(\ref{gvphi}) along with the $\mathcal{U}_0(\Phi)$ 
potential added to $\Omega_{\mbox{\tiny PNJL}}(\rho,0)$ lead to the following equations for the 
pressure and energy density, respectively,
\begin{align}
P_{\mbox{\tiny PNJL0}}&(\rho) = -\mathcal{U}_{\mbox{\tiny PNJL0}}(\rho,\rho_s,\Phi) + 
G_V\rho^2-G_s\rho^2_{s} 
\nonumber\\
&+\frac{\gamma}{2\pi^2}\int^{\Lambda}_{0} dk 
k^2(k^2+M^2)^{1/2} + \mathcal{E}_{\rm vac}
\nonumber\\
&+ \frac{\gamma}{6\pi^2}\int_{0}^{k_F}dk  
\frac{k^4}{(k^2+M^2)^{1/2}}  = -\Omega_{\mbox{\tiny PNJL0}}(\rho)
\label{presspnjl0} 
\end{align}
and
\begin{align}
\mathcal{E}_{\mbox{\tiny PNJL0}}(\rho) &= \mathcal{U}_{\mbox{\tiny PNJL0}}(\rho,\rho_s,\Phi) - 
2G_V\Phi^2\rho^2 + G_V\rho^2 + G_s\rho^2_{s} 
\nonumber\\
&- \frac{\gamma}{2\pi^2}\int_{k_F}^{\Lambda}dk 
k^2(k^2+M^2)^{1/2} - \mathcal{E}_{\rm vac},
\label{depnjl0}
\end{align}
in which it was possible to define a Polyakov potential as being
\begin{align}
\mathcal{U}_{\mbox{\tiny PNJL0}}(\rho_s,\rho,\Phi) = G_V\Phi^2\rho^2 - G_s\Phi^2\rho_s^2 
+ \mathcal{U}_0(\Phi)
\label{upnjl0}
\end{align}
which includes quarks as source of $\Phi$, as one should expect from QCD where quarks are also 
sources for the gluon field.

The model built above, is named as PNJL0 model, and it has  constituent quark mass and quark chemical 
potential given by:
\begin{eqnarray}
M = m - 2G_s(1-\Phi^2)\rho_s~,
\label{mpnjl0}
\end{eqnarray}  
and
\begin{eqnarray}
\mu = 2G_V(1 - \Phi^2) \rho + (k_F^2+M^2)^{1/2},
\label{mupnjl0}
\end{eqnarray}
respectively. Furthermore, the quark density is written in terms of the quark Fermi momentum as 
$\rho=(\gamma/6\pi^2)k_F^3$, and the quark condensate reads
\begin{eqnarray}
\rho_s=-\frac{\gamma M}{2\pi^2}\int_{k_F}^\Lambda dk\frac{k^2}{(k^2+M^2)^{1/2}},
\label{rhospnjl0}
\end{eqnarray}
which contributes to the scalar quark density.

It is worthwhile to notice that Eqs.~(\ref{presspnjl0}) and~(\ref{depnjl0}) can be seen as the 
$T\rightarrow 0$ limit of Eqs.~(\ref{ppnjl}) and~(\ref{depnjl}). Another important aspect in 
defining the Polyakov potential $\mathcal{U}_{\mbox{\tiny PNJL0}}(\rho_s,\rho,\Phi)$ is the 
presence of the backreaction of quarks in the gluons sector, as we have already pointed out. 
The inverse backreaction, namely, gluons  affecting the quark sector is already intrinsic 
in the original PNJL model, see for instance, the 
generalized Fermi-Dirac distribution functions in Eqs.~(\ref{fdmp}) and~(\ref{fdmap}). In the PNJL0 
model the backreaction is complete (each sector interacts each other): 
 the effective quark interactions vanishes at the deconfinement 
phase and  $\mathcal{U}_0(\Phi)$ is included in the grand canonical potential
to assure confinement physics at $T=0$.

Just to be complete, we should mention that another way of 
including quark effects in the gluon sector was given, for example, in 
Refs.~\cite{herbst,schaefer}, in which the authors impose a $N_f$ and $\mu$ dependence on $T_0$ 
in the Polyakov potentials, namely, $T_0 \rightarrow T_0(N_f,\mu)$.

\subsection{$\Phi\ne0$ solutions for the PNJL0 model}

The inclusion of $\mathcal{U}_0(\Phi)$ in the Polyakov potential given by Eq.~(\ref{upnjl0}) 
enables the PNJL0 model to present $\Phi\ne0$ solutions for the condition 
$\partial\Omega_{\mbox{\tiny PNJL0}}/\partial\Phi=0$ at zero temperature. Therefore, it becomes 
possible the study of the deconfinement dynamics in the $T=0$ regime, with the dimensionless 
constant $a_3$ in Eq.~(\ref{u0}) regulating this effect.

\begin{figure}[!htb] 
\centering
\includegraphics[scale=0.33]{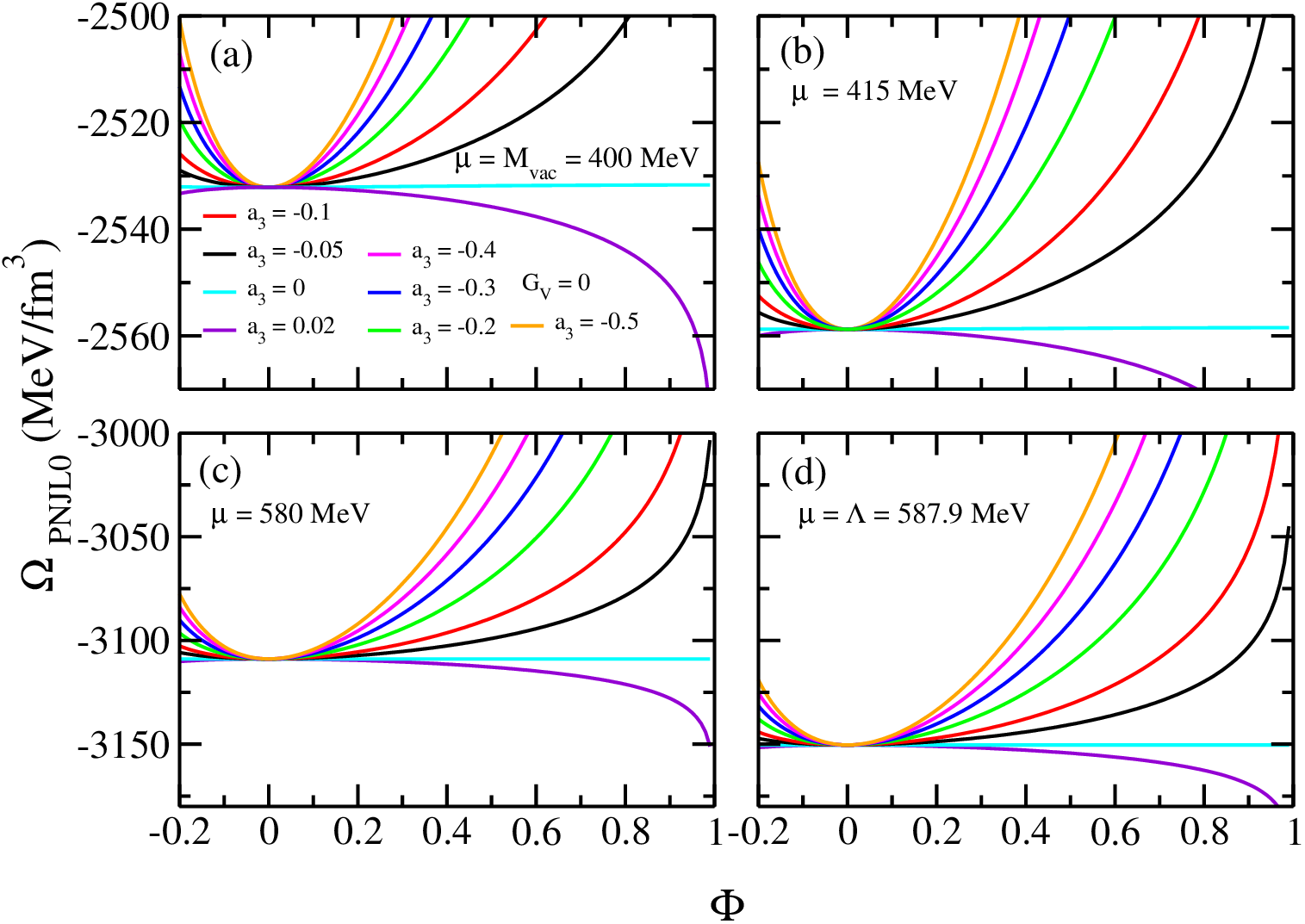}
\caption{$\Omega_{\mbox{\tiny PNJL0}}\times\Phi$ for $G_V=0$ with different values of $a_3$. 
For each panel the quark chemical potential is given by (a)~$\mu=M_{\mbox{\tiny vac}}=400$~MeV, 
(b)~$\mu=415$~MeV, (c)~$\mu=580$~MeV, and (d)~$\mu=\Lambda=587.9$~MeV.} 
\label{omegaphigv0}
\end{figure}

We investigate how $\Omega_{\mbox{\tiny PNJL0}}$ behaves as 
a function of $\Phi$ firstly for $G_V=0$. In Fig.~\ref{omegaphigv0} we display this thermodynamical 
potential, obtained from $\Omega_{\mbox{\tiny PNJL0}}=-P_{\mbox{\tiny PNJL0}}$, 
Eq.~(\ref{presspnjl0}), with the self-consistent equations (\ref{mpnjl0}) and~(\ref{rhospnjl0}) 
implemented, and without the condition $\partial\Omega_{\mbox{\tiny PNJL0}}/\partial\Phi=0$ taken 
into account. We also use a parametrization of Ref.~\cite{buballa}, namely, $\Lambda=587.9$~MeV, 
$G_s\Lambda^2=2.44$ and $m=5.6$~MeV for this case and all the other ones. Such values produce 
$M_{\mbox{\tiny vac}}=400$~MeV and $\langle\overline{u}u\rangle^{1/3}_{\mbox{\tiny 
vac}}=-240.8$~MeV for the vacuum values,  $m_\pi=135$~MeV and $f_\pi=92.4$~MeV for the pion mass, 
and decay constant, respectively.

\begin{figure}[!htb] 
\centering
\includegraphics[scale=0.33]{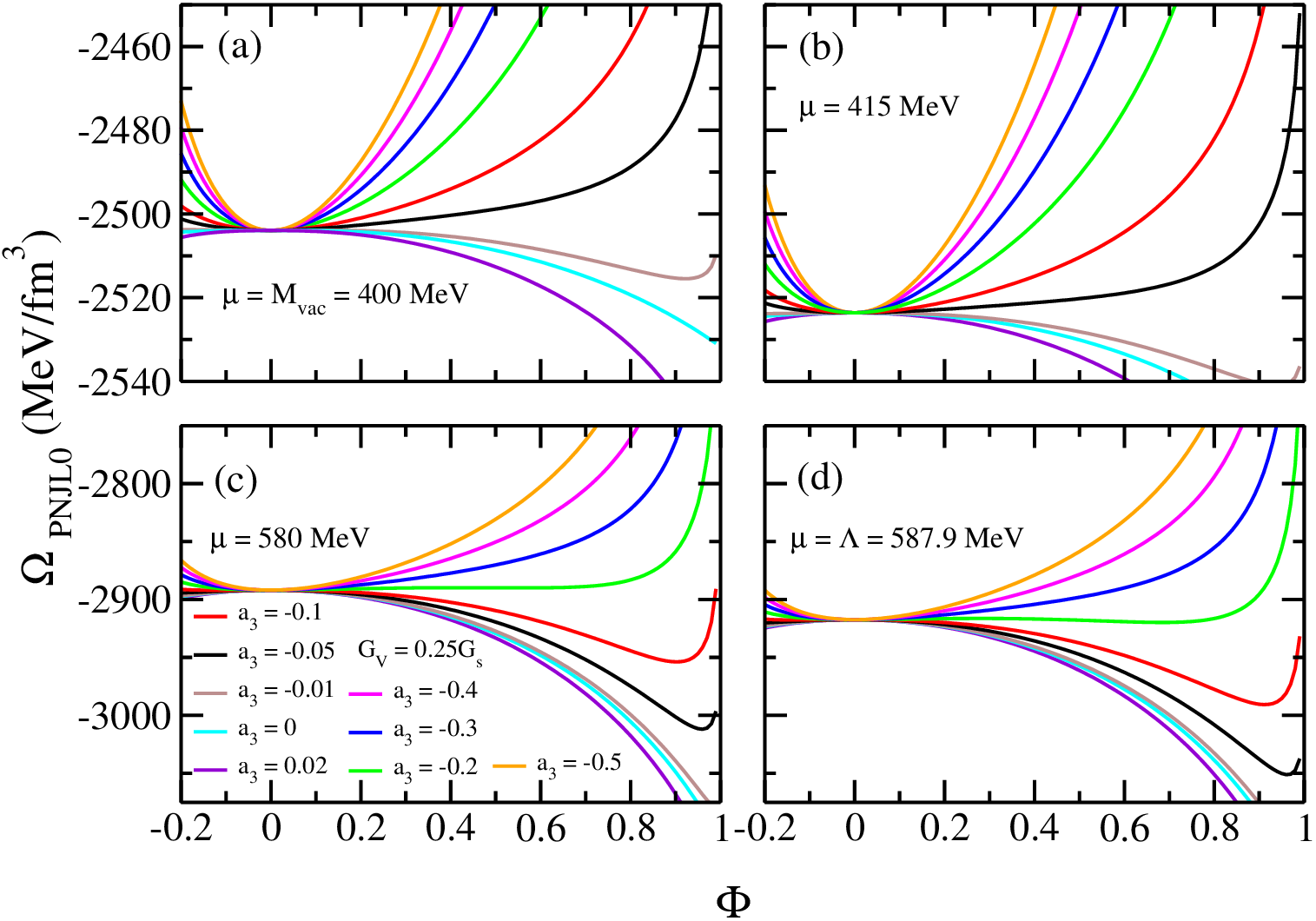}
\caption{The same as in Fig.~\ref{omegaphigv0}, but for $G_V=0.25G_s$.} 
\label{omegaphigv025}
\end{figure}

Some features can be observed from the results depicted in Fig.~\ref{omegaphigv0}. The first one is 
that the variation of~$a_3$ does not produce any change in the value of $\Phi$ concerning the 
minimum of $\Omega_{\mbox{\tiny PNJL0}}$. For all $a_3$ values one has $\partial\Omega_{\mbox{\tiny 
PNJL0}}/\partial\Phi=0$ only at $\Phi=0$. Notice also that positive~$a_3$ values induce 
$\Omega_{\mbox{\tiny PNJL0}}$ to change its concavity, i.e, a non physical configuration. This 
effect is also verified for other $\mu$ values different from those shown in Fig.~\ref{omegaphigv0}. 
Even for the extreme value of $\mu=\Lambda$, Fig.~\ref{omegaphigv0}{\color{blue}d}, there is no 
indication that the system presents $\Phi\neq 0$. Notice that the variation of $\mu$ for a fixed 
$a_3$ is also not enough to produce $\Phi\neq 0$ solutions for 
$\partial\Omega_{\mbox{\tiny PNJL0}}/\partial\Phi=0$. The increase in $\mu$ 
decreases  $\Omega_{\mbox{\tiny PNJL0}}(\Phi=0)$ as the only effect. This lack of $\Phi\neq 0$ 
solutions means that the model can not bring confinement effects for the analyzed region of the 
quark chemical potential, namely, $M_{\mbox{\tiny vac}}\leqslant\mu\leqslant\Lambda$. Therefore, for 
this case the PNJL0 model reduces to the NJL one  as the PNJL model does. However, this 
picture is radically modified when $G_V\ne 0$ as one can see in Fig.~\ref{omegaphigv025}, where
$G_V=0.25G_s$ was used.

The results displayed in Fig.~\ref{omegaphigv025} show that for $G_V=0.25G_s$ there is a global 
minima for $\Omega_{\mbox{\tiny PNJL0}}$ at $\Phi\neq 0$. As an example, see that for 
$\mu=\Lambda=587.9$~MeV, Fig.~\ref{omegaphigv025}{\color{blue}d}, this minimum is found at 
$\Phi\approx 0.9$ for $a_3=-0.1$. For the same $a_3$ value, $\Phi=0$ is the unique global minimum 
for $\mu=M_{\mbox{\tiny vac}}=400$~MeV. This means that there is an intermediate value of $\mu$ in 
which there is two minima for $\Omega_{\mbox{\tiny PNJL0}}$, namely, one of them at $\Phi=0$ and the 
other one at $\Phi\neq 0$. Physically, this means that there exists a certain $\mu$ value, for the 
$a_3=-0.1$ case, in which the transition from a quark confined phase ($\Phi=0$) to a deconfined 
one ($\Phi\neq0$) takes place. For the PNJL0 model, this $\Phi\ne 0$ solutions are found for 
$G_V\ne 0$, i.e., the repulsive vector channel plays an important role for the emergence of 
deconfinement effects at zero temperature regime.

\subsection{Confinement/deconfinement phase transition}

\begin{figure}[!htb]
\centering
\includegraphics[scale=0.3]{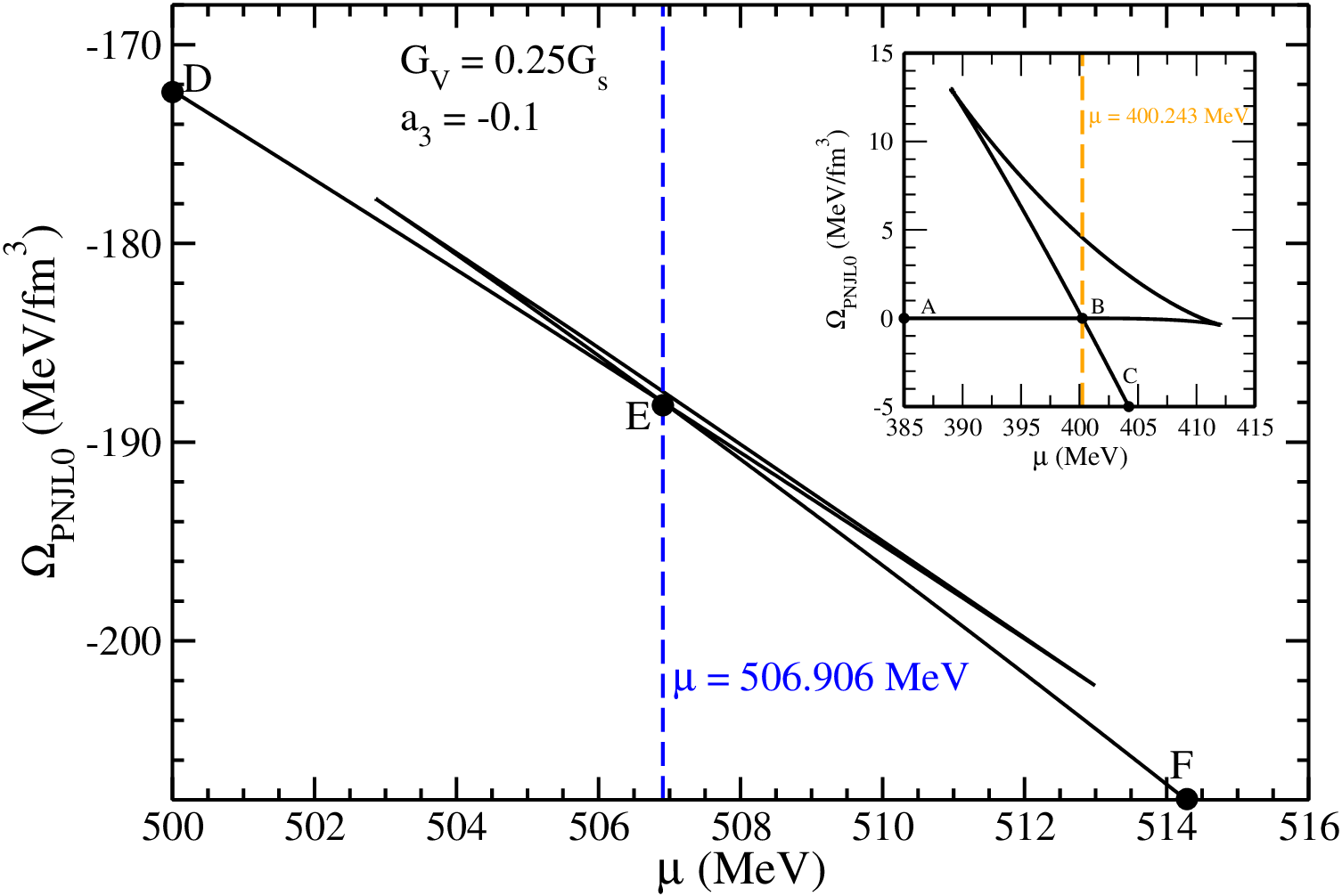}
\caption{$\Omega_{\mbox{\tiny PNJL0}}$ as a function of the quark chemical potential~($\mu$) for 
$G_V=0.25G_s$ and $a_3=-0.1$.}
\label{omegamu1}
\end{figure}

In order to correctly identify the quark confined and deconfined thermodynamical phases in the 
PNJL0 model, we compute the grand canonical potential given by Eq.~(\ref{presspnjl0}) with the 
self-consistent equations (\ref{mpnjl0}) and~(\ref{rhospnjl0}) implemented, but now along with the 
condition $\partial\Omega_{\mbox{\tiny PNJL0}}/\partial\Phi=0$. The result of this calculation, for 
$G_V=0.25G_s$ and $a_3=-0.1$, is presented in Fig.~\ref{omegamu1}.
This figure displays the typical feature of systems that present first order phase transition, 
namely, non unique values for the thermodynamical potential that describes the system 
($\Omega_{\mbox{\tiny PNJL0}}$) as a function of the intensive quantity ($\mu$). Thermodynamical 
stability~\cite{callen} requires that in the range of $500\mbox{ 
MeV}\leqslant\mu\lesssim 514\mbox{ MeV}$ the dependence of the grand canonical potential with $\mu$ 
is the one defined by the $DEF$ curve. The $E$
point indicates the chemical potential value at which  a first order phase transition takes place, 
namely, a confinement/deconfinement one, with $\Phi$ the order parameter in this case as we will
discuss later on. We name the chemical potential at this point as $\mu_{\mbox{\tiny 
conf}}$ with the value of $\mu_{\mbox{\tiny conf}} = 506.906$~MeV. 

\begin{figure}[!htb]
\centering
\includegraphics[scale=0.3]{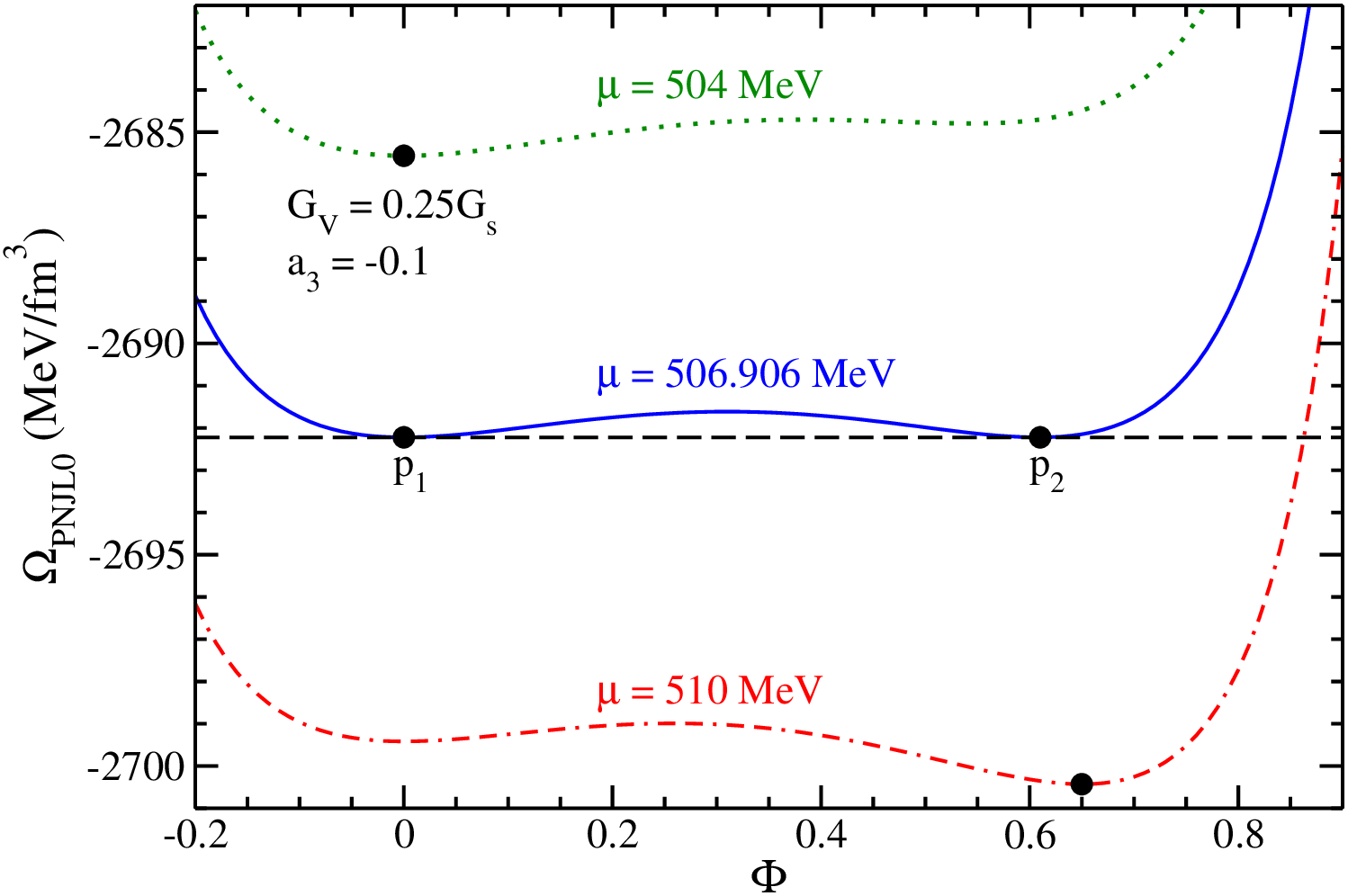}
\caption{$\Omega_{\mbox{\tiny PNJL0}}$ as a function of~$\Phi$ for different $\mu$ values. Curves 
constructed for $G_V=0.25G_s$ and $a_3=-0.1$.}
\label{omegaphi}
\end{figure}

Another way to determine $\mu_{\mbox{\tiny conf}}$ is from the analysis of $\Omega_{\mbox{\tiny 
PNJL0}}$ as a function of~$\Phi$ without imposing the condition 
$\partial\Omega_{\mbox{\tiny PNJL0}}/\partial\Phi=0$, for each fixed value of $\mu$. 
The different grand potentials obtained for each $\mu$  are depicted 
in~Fig.~\ref{omegaphi}.
The value of $\mu_{\mbox{\tiny conf}}$ is obtained when two minima of the thermodynamical 
potential appear for distinct values of $\Phi$ with the same  $\Omega_{\mbox{\tiny 
PNJL0}}$, see points $p_1$ and $p_2$ in the $\mu=\mu_{\mbox{\tiny conf}}=506.906$~MeV curve. The 
$\Phi$ values associated to these points delimit the confined quark phase (point $p_1$, $\Phi=0$) 
and the deconfinement one (point $p_2$, $\Phi\neq 0$). For curves where $\mu\neq\mu_{\mbox{\tiny 
conf}}$, there is only one global minimum in $\Omega_{\mbox{\tiny PNJL0}}$. For such cases, the 
system is exclusively in one of the two thermodynamic phases concerning the quark confinement. 

\begin{figure}[!htb] 
\centering
\includegraphics[scale=0.3]{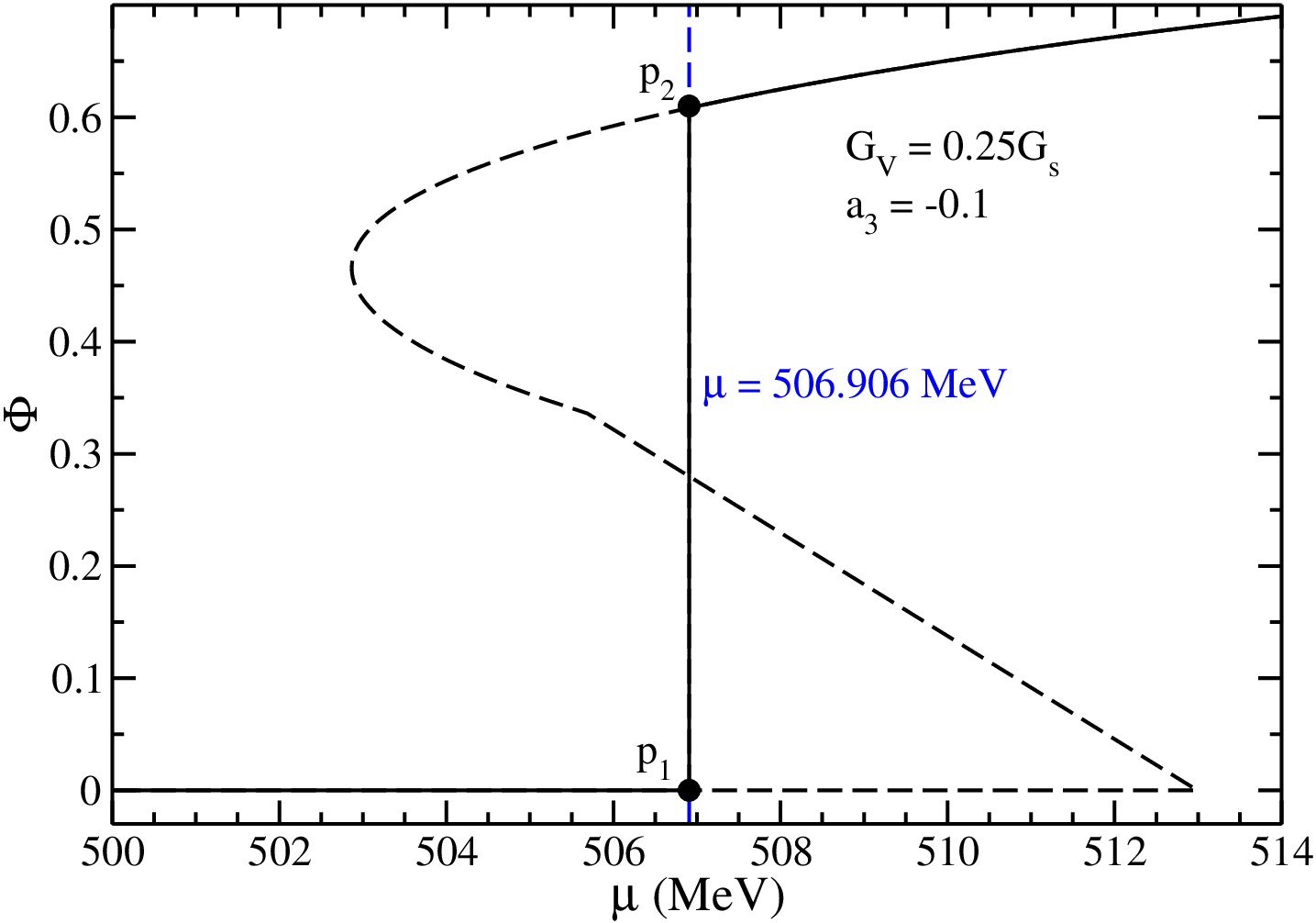}
\caption{$\Phi$ as a function of $\mu$ for the PNJL0 model with $G_V=0.25G_s$ and $a_3=-0.1$.}
\label{phimu}
\end{figure}

In Fig.~\ref{omegaphi} we also notice that the $\mu=504$~MeV curve is in a confined phase, since 
the minimum is attained at $\Phi=0$, but for $\mu=510$~MeV, a $\Phi\neq 0$ value is the possible 
one and identifies the system in deconfined phase. Only at $\mu=\mu_{\mbox{\tiny conf}}$ the system 
undergoes a first order phase transition. Such procedure of searching for two global minima in the 
thermodynamical potential was also used, for instance, in the analysis of mean-field hadronic 
models~\cite{delfino}, as well as for the NJL model at finite temperature~\cite{yazaki}. Both models 
present the same kind of first order phase transition, but in different environments and with 
different order parameters. In the case of the PNJL0 model at zero temperature, $\Phi$ is the order 
parameter related to the confined/deconfined phase transition. Its dependence with $\mu$, obtained 
through $\partial\Omega_{\mbox{\tiny PNJL0}}/\partial\Phi=0$, is shown in Fig.~\ref{phimu}.
The equilibrium $\Phi$ dependence with $\mu$ is the one defined by the full line. The position of 
the jump in $\Phi$ is determined by $\mu=\mu_{\mbox{\tiny conf}}$, found by the aforementioned 
method of searching for two global minima in $\Omega_{\mbox{\tiny PNJL0}}$. The dashed line 
corresponds to the eliminated branches of $\Omega_{\mbox{\tiny PNJL0}}$ in Fig.~\ref{omegamu1}.

\subsection{Quarkyonic phase and effects of $a_3$ and $G_V$ on the PNJL0 model}

Another thermodynamical phase structure is also observed at lower $\mu$ values, namely, $385\mbox{ 
MeV}\leqslant\mu\leqslant 415\mbox{ MeV}$, as pointed out by the inset of Fig.~\ref{omegamu1}. In 
that region, it is verified that the correct $\Omega_{\mbox{\tiny PNJL0}}\times\mu$ curve must be 
the one described by the $ABC$ line. The first order phase transition is given by the transition of 
the system from a broken chiral symmetry region to a restored one (the quark condensate is the order 
parameter in this case). By naming the chemical potential at the $B$ point as $\mu_{\mbox{\tiny 
chiral}}$, we obtain the value of $400.243$~MeV. Notice that as $\Phi=0$ in this region, the same 
thermodynamical structure is also present in the NJL model. In this case, Eqs.~(\ref{presspnjl0}) 
to~(\ref{mupnjl0}) are reduced to the NJL ones for $\Phi=0$ (confined phase of the PNJL0 model). 
Strictly speaking, the PNJL0 model is exactly the NJL one until $\mu=\mu_{\mbox{\tiny conf}}$, 
where 
nonzero $\Phi$ values occur, i.e., in our approach the NJL thermodynamical phases can be seen as 
contained in the PNJL0 model. 

\begin{figure}[!htb]
\centering
\includegraphics[scale=0.3]{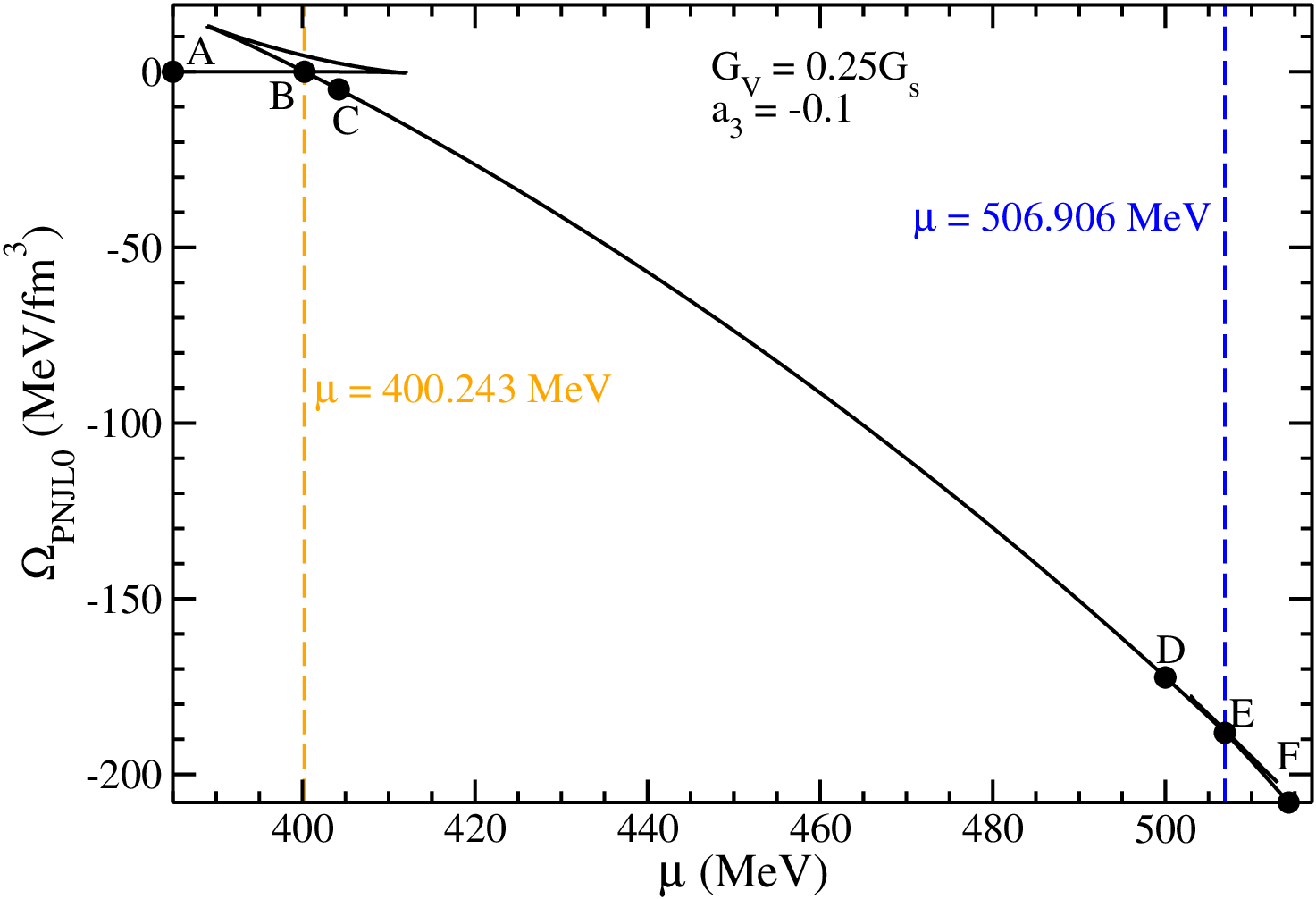}
\caption{The same as in Fig.~\ref{omegamu1}, but for a larger $\mu$ region.}
\label{omegamu2}
\end{figure}

A wider picture of the PNJL0 model, encompassing the two phase 
transition regions, is shown in Fig.~\ref{omegamu2}. The thermodynamical stable curve in this case 
is the one described by the $ABCDEF$ line.

\begin{figure}[!htb] 
\centering
\includegraphics[scale=0.3]{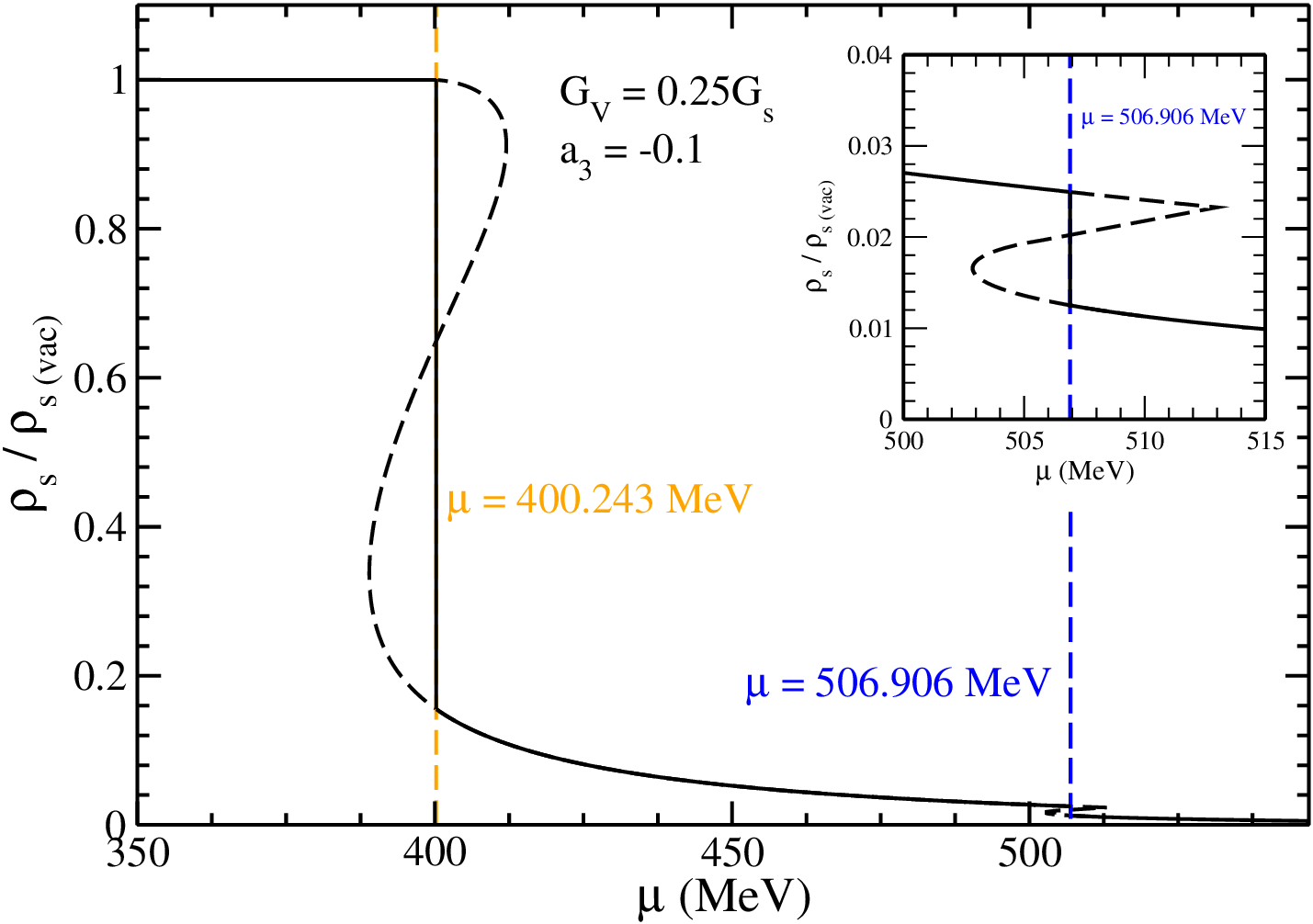}
\caption{Chiral condensate, in units of its vacuum value, as a function of $\mu$ for the PNJL0 model 
with $G_V=0.25G_s$ and $a_3=-0.1$.}
\label{rhosmu}
\end{figure}

In Fig.~\ref{rhosmu}, we show the $\mu$ dependence of the chiral condensate for the PNJL0 model.
For chemical potential values smaller than $\mu_{\mbox{\tiny chiral}}=400.243$~MeV, the model 
behaves as the NJL one, as already discussed. Exactly at $\mu=\mu_{\mbox{\tiny conf}}=506.906$~MeV, 
the discontinuity in $\Phi$ also affects $\rho_s$ due to the backreaction mechanism presented in 
the PNJL0 model. We remark in the inset of the Fig.~\ref{rhosmu} the discontinuity induced in 
$\rho_s$ due to the one observed in~$\Phi$ (Fig.~\ref{phimu}). The stable curve for $\rho_s$ is the 
one described by the full line.

Fig.~\ref{rhosmu} is also useful to identify another phase of the strongly interacting matter, 
namely, the one defined in the region of $\mu_{\mbox{\tiny chiral}} \leqslant \mu \leqslant 
\mu_{\mbox{\tiny conf}}$. In this range, quark matter is chirally symmetric but still confined since 
the quark condensate is nearly vanishing and the traced Polyakov loop is zero. Only at 
$\mu\geqslant\mu_{\mbox{\tiny conf}}$, the deconfined quark phase is reached, i.e, one has 
$\Phi\neq 0$. This specific $\mu$ region in the range of $\mu_{\mbox{\tiny chiral}}\leqslant\mu 
\leqslant \mu_{\mbox{\tiny conf}}$ is identified as the quarkyonic phase, in the notation of 
Refs.~\cite{fukushima3,abuki,mcnpa09,buisseret,mcnpa07,mcnpa08}, for instance. The emergence of 
this phase is not possible in the usual NJL model since there is no information regarding 
confinement effects as there is in the PNJL0 model.

\begin{figure}[!htb] 
\centering
\includegraphics[scale=0.3]{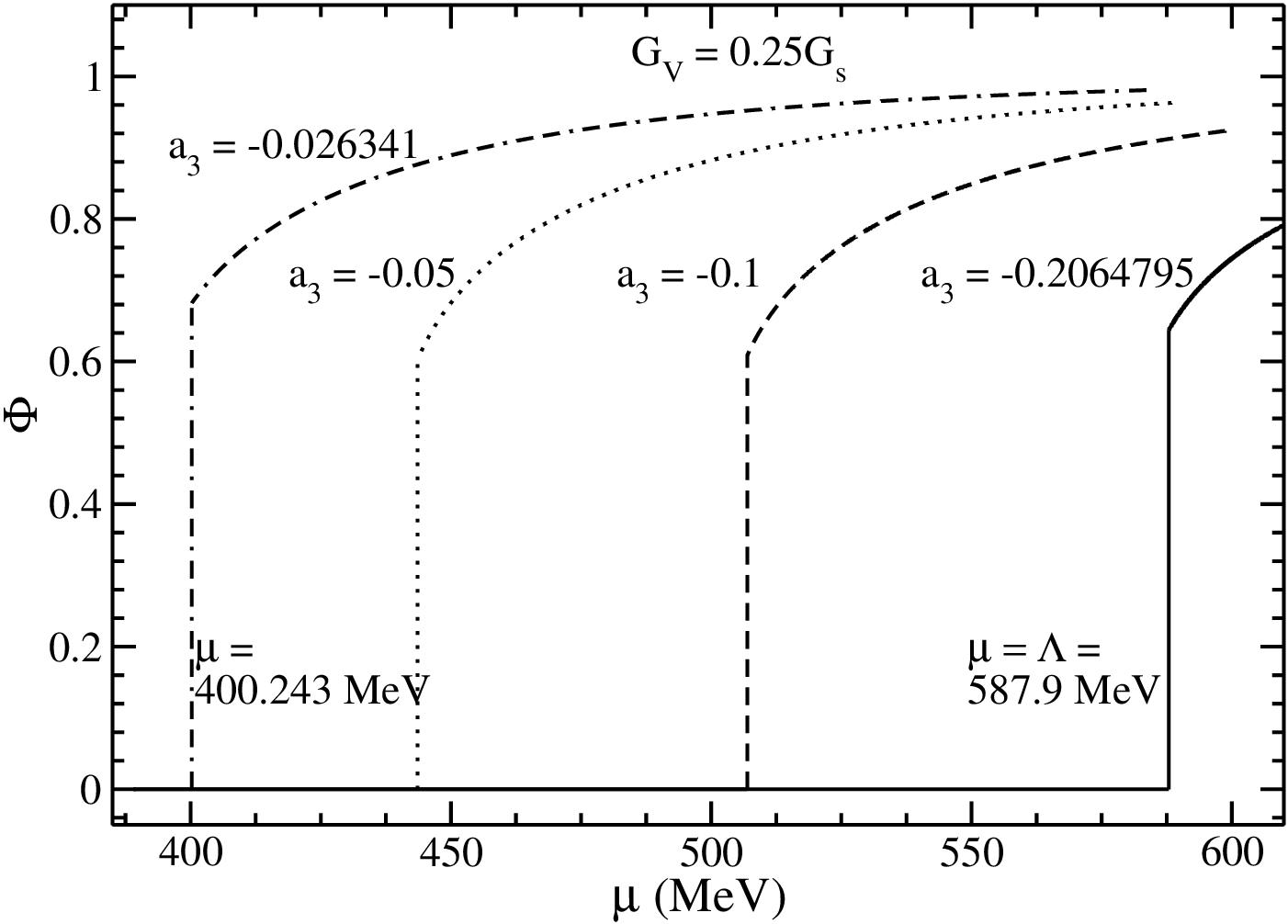}
\caption{$\Phi\times\mu$ for the PNJL0 model with $G_V=0.25G_s$ and different $a_3$ values.}
\label{phimua3}
\end{figure}

\begin{figure}[!htb] 
\centering
\includegraphics[scale=0.3]{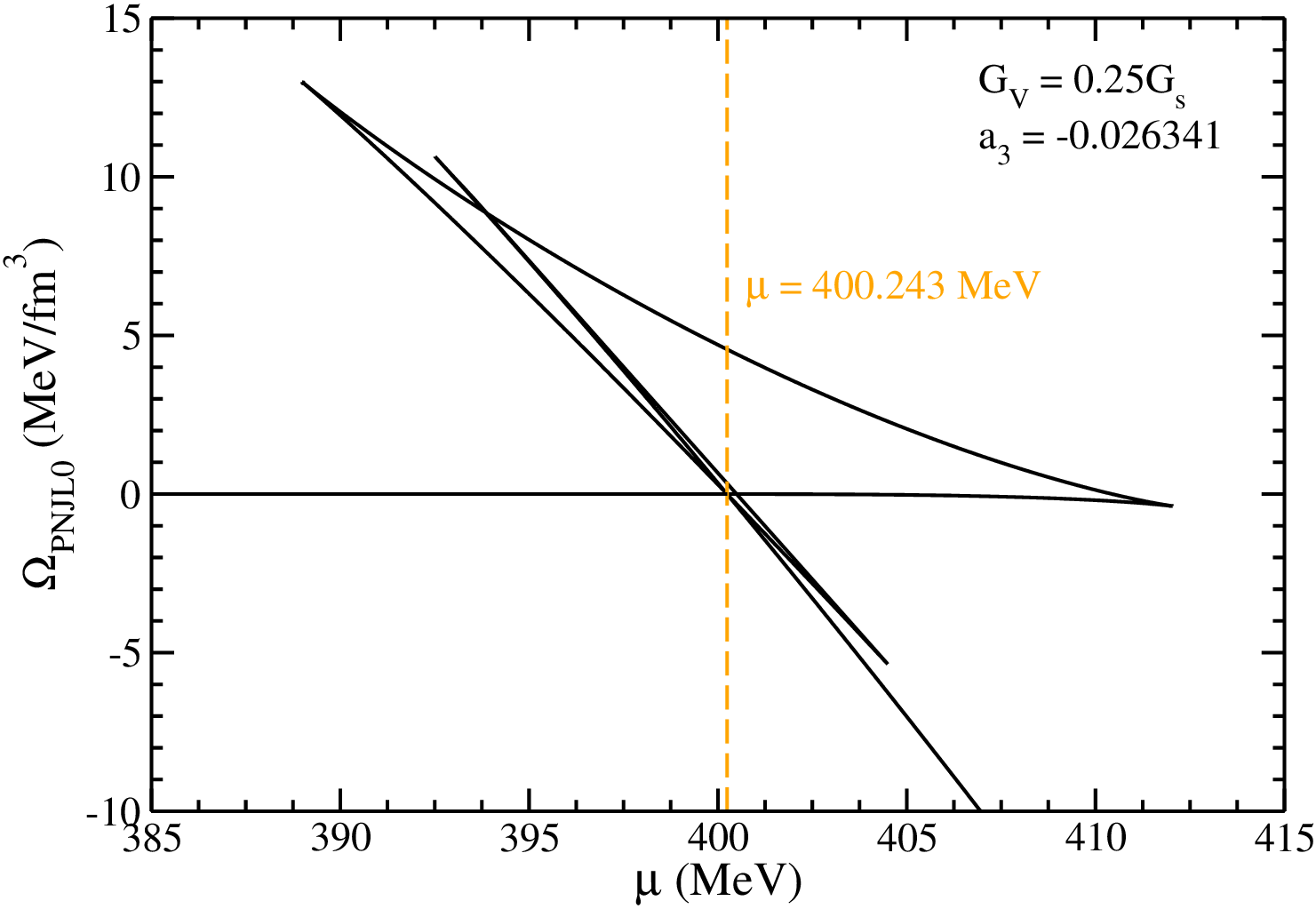}
\caption{$\Omega_{\mbox{\tiny PNJL0}}$ as a function of the quark chemical potential~($\mu$) for 
$G_V=0.25G_s$ and $a_3=-0.026341$.}
\label{omegamu3}
\end{figure}

We also investigate the effects of the variation of the  $a_3$ and $G_V$ parameters in the PNJL0 
model. In Fig.~\ref{phimua3} we show $\Phi$ as a function of $\mu$ for different $a_3$ values chosen 
in order to produce $\mu_{\mbox{\tiny conf}}=\mu_{\mbox{\tiny chiral}}$ and $\mu_{\mbox{\tiny 
conf}}=\Lambda$ for the chemical potentials related to the two phase transitions (broken/restored 
chiral symmetry phase transition and confinement/deconfinement one).
Note that the effect of the~$a_3$ increasing is to shrink the quarkyonic phase until its complete 
elimination, in this case for $a_3\sim -0.026$. For this particular value of~$a_3$, 
$\Omega_{\mbox{\tiny PNJL0}}\times\mu$ present two crossing points exactly at the same~$\mu$, 
namely, $\mu=\mu_{\mbox{\tiny chiral}}=\mu_{\mbox{\tiny conf}}=400.243$~MeV, as we see in 
Fig.~\ref{omegamu3}.

\begin{figure}[!htb] 
\centering
\includegraphics[scale=0.3]{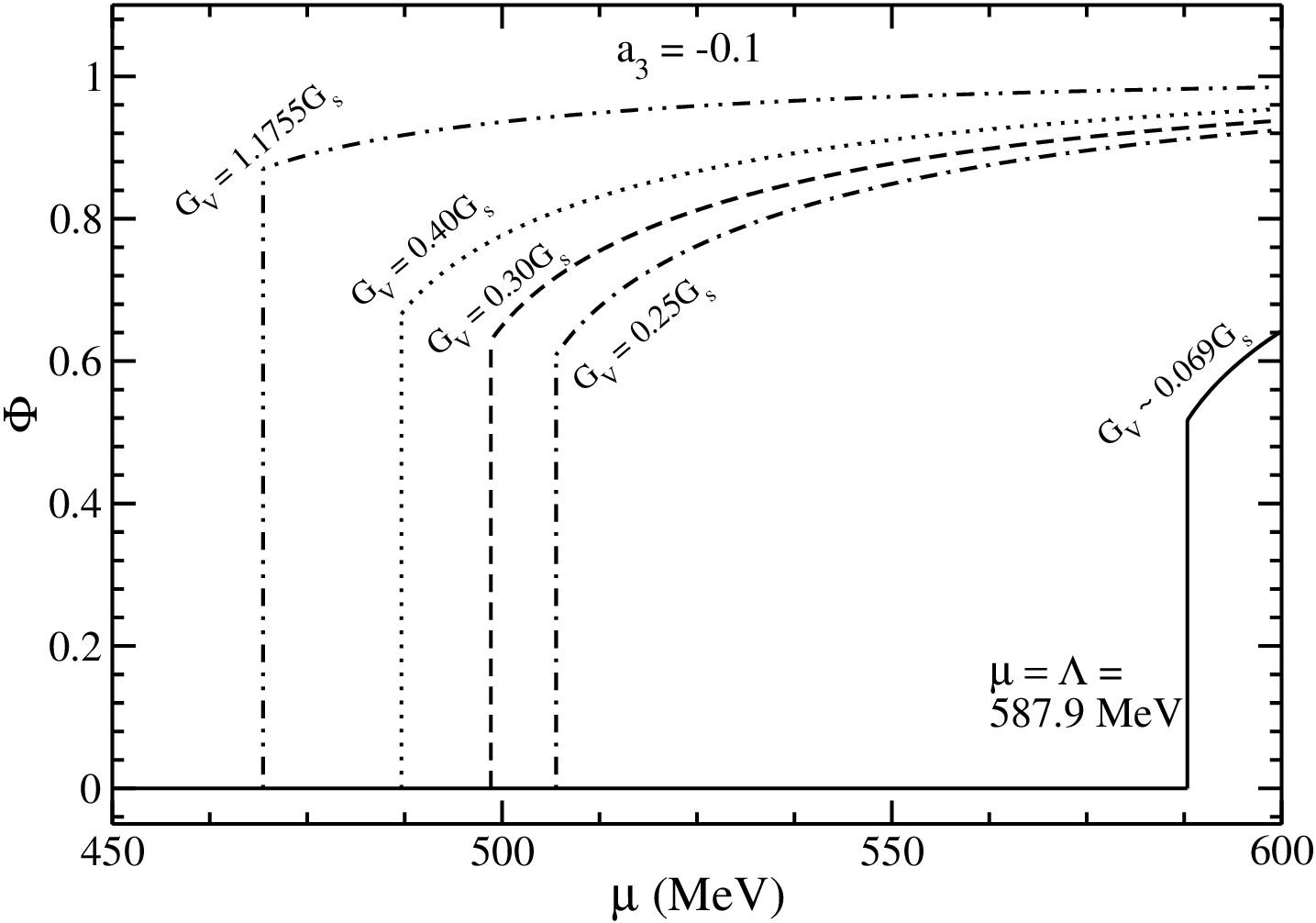}
\caption{$\Phi$ as a function of the quark chemical potential for $a_3=-0.1$ and different 
$G_V$ values. We also remark that as $G_V$ increases above $\sim G_s$, the lines begin 
to move significantly to the right. The reader can confer this effect in Fig.~\ref{phimugv2}.}
\label{phimugv1} 
\end{figure}

Finally, we study the impact of the $G_V$ variation in the model. In Fig.~\ref{phimugv1} it is 
depicted how the traced Polyakov loop is affected by the strength of the vector channel interaction. 
It can be seen that the increase of $G_V$ also moves the entire $\Phi$ curve to the 
direction of lower $\mu$ values. As a direct consequence, the quarkyonic region begins to decrease 
in size (as $G_V$ increases) until the situation in which $G_V\sim 1.18G_s$. In this case, the 
quantity defined as $\Delta\mu=\mu_{\mbox{\tiny conf}}-\mu_{\mbox{\tiny chiral}}$ is vanishing. 
Thereafter, $\mu_{\mbox{\tiny chiral}}$ becomes greater than $\mu_{\mbox{\tiny conf}}$, indicating 
that chiral symmetry restoration takes place after deconfinement. The change of $\mu_{\mbox{\tiny 
chiral}}$ as a function of $G_V$ is an effect observed also in the original NJL 
model~\cite{buballa}. Actually, the increase of~$G_V$ moves the point of broken/restored chiral 
symmetry phase transition to higher $\mu$ values, making possible the PNJL0 model to present 
$\mu_{\mbox{\tiny chiral}}>\mu_{\mbox{\tiny conf}}$. In order to avoid this situation, we restrict 
$G_V$ to the maximum value that leads to $\Delta\mu=0$, namely, 
$G_V\sim 1.18G_s$.\footnote{For $G_V$ values that eliminate the first order phase 
transition, we calculate $\mu_{\mbox{\tiny chiral}}$ as the chemical potential related to the peak 
of~$|\frac{\partial \rho_s}{\partial\mu}|$.}

Studies with the aim of correctly limit the $G_V$ values were performed, for 
instance, in Refs.~\cite{carignano,kashiwa,rapp} where the range of $0.25G_s\leqslant 
G_V\leqslant0.50G_s$ was found. In Refs.~\cite{nosso2,bratovic}, on the other hand, the more broad 
range of $0.30G_s\leqslant G_V\leqslant3.2G_s$ was used. Here, it is possible to adopt a criterion 
in order to determine a range of $G_V$ based on the results shown for the PNJL0 model, namely, we 
define $G_V$ inside an interval of $ G_V^{\mbox{\tiny min}}\leqslant G_V\leqslant G^{\mbox{\tiny 
max}}_V $, where $G^{\mbox{\tiny min}}_V $ is the value that produces $\mu_{\mbox{\tiny 
conf}}=\Lambda$, and $G_V^{\mbox{\tiny max}}$ is the value that leads to 
$\mu_{\mbox{\tiny conf}}=\mu_{\mbox{\tiny chiral}}$~\mbox{$(\Delta\mu=0)$}, according to the 
previous discussion. This generates a quarkyonic phase starting at $\mu=\mu_{\mbox{\tiny chiral}}$ 
and extending up to a certain typical energy scale of the system as $\Lambda$, for instance. 
Such an approach also avoids a confinement/deconfinement phase transition 
taking place before the broken/restored chiral symmetry one. In the specific case of $a_3=-0.1$, 
this criterion leads to $G^{\mbox{\tiny min}}_V\sim 0.069G_s$ and $G^{\mbox{\tiny 
max}}_V\sim 1.18G_s$. 

\begin{figure}[!htb] 
\centering
\includegraphics[scale=0.29]{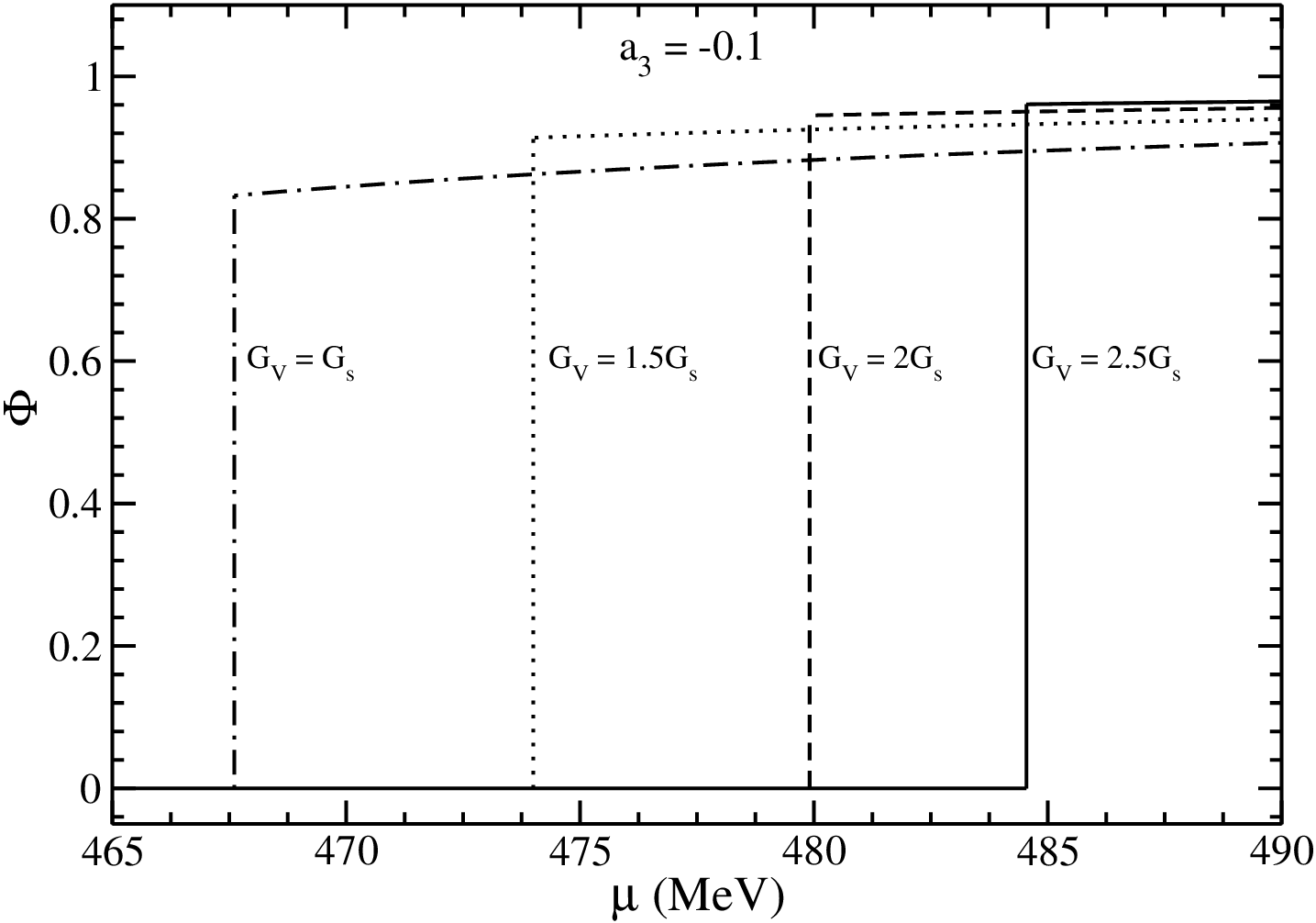}
\caption{$\Phi$ as a function of $\mu$ for $G_V/G_s$ = 1, 1.5, 2 and 2.5.}
\label{phimugv2}
\end{figure}

In Fig.~\ref{phimugv2} we also show the evolution of $\Phi\times\mu$ curves for higher~$G_V$ 
values. Notice that the mainly effect in these cases is to move the $\Phi$ curve to the increasing 
$\mu$ direction.

As a last comment, we reinforce that the PNJL0 model studied in this work (zero temperature 
regime) produces first order phase transitions for the traced Polyakov loop, as a function of the 
quark chemical potential, for different values of the free parameters $G_V$ and $a_3$, as observed 
in Figs.~\ref{phimu}, \ref{phimua3}, \ref{phimugv1}, and~\ref{phimugv2}. The same kind of phase 
transition is also found in Refs.~\cite{Schramm1,Schramm2}, where the Polyakov potential given in 
Eq.~(\ref{ds10}) was implemented in a hybrid SU(3) chiral model that contains both hadrons and 
quarks as degrees of freedom. Besides that study, in Ref.~\cite{pnjl0outro} another version of the 
PNJL model at $T=0$ was proposed. In that model, a quark density dependence is introduced in the 
$b_2(T)$ function of the Polyakov potential given in Eq.~(\ref{rrw06}). A signal of first order 
phase transition is also verified in their $\Phi\times\mu$ curves, but with the discontinuous jump of
the trace Polyakov loop coinciding with the hadron-quark phase transition.

\section{Summary and concluding remarks}
\label{secsummary}

In this work we extend a previous study performed in Ref.~\cite{mattos} in which a version of the
Polyakov-Nambu-Jona-Lasinio model at zero temperature (PNJL0 model) was proposed. Here we 
explicitly show that it is possible to preserve the confinement effects of the model even at 
$T=0$ by imposing a $\Phi$ dependence in the strengths of the scalar and vector interaction 
channels, see Eqs.~(\ref{gsphi}) and~(\ref{gvphi}). This modification leads to a Polyakov potential 
given by 
\begin{align}
\mathcal{U}_{\mbox{\tiny PNJL0}}(\rho_s,\rho,\Phi) &= G_V\Phi^2\rho^2 - G_s\Phi^2\rho_s^2
\nonumber\\ 
&+ a_3T_0^4\mbox{ln}(1-6\Phi^2+8\Phi^3-3\Phi^4),
\end{align}
containing the backreaction of the quarks in the gluonic sector, but also the influence of the 
latter on the former. {The last term in $\mathcal{U}_{\mbox{\tiny PNJL0}}$ ensures nonvanishing 
solutions for $\Phi$ and also limits the $\Phi$ values to $1$}. We show that $\Omega_{\mbox{\tiny 
PNJL0}}\times\Phi$ present $\Phi\ne0$ solutions only for $G_V\neq0$. Therefore, the vector channel 
plays an important role in the PNJL0 model since it ensures the possibility of observing the 
confinement effects represented by nonvanishing traced Polyakov loop values. We show how these 
solutions generates first order phase transitions related to the confined/deconfined quark phases, 
since $\Phi$ present a discontinuous 
jump as a function of the quark chemical potential ($\Phi$ is the order parameter). The signature of 
this phase transition is identified in the $\Omega_{\mbox{\tiny PNJL0}}\times\mu$ curve where a crossing 
point is observed, see Fig.~\ref{omegamu1}. Thermodynamical stability establishes that it is also 
possible to identify such a transition through the search of the chemical potential that produces 
two global minima, with the same $\Omega_{\mbox{\tiny PNJL0}}$ value, in the $\Omega_{\mbox{\tiny 
PNJL0}}\times\Phi$ curve, see Fig.~\ref{omegaphi}. 

We also show that the first order phase transition related to the broken/restored chiral symmetry 
is still present in the PNJL0 at the same value of $\mu$ as in the original NJL model, as one can 
see in Fig.~\ref{omegamu2}. The first crossing point indicates this transition, with the quark 
condensate being the order parameter. The region $\mu_{\mbox{\tiny chiral}} \leqslant \mu \leqslant 
\mu_{\mbox{\tiny conf}}$ is identified as the quarkyonic phase, namely, chirally symmetric 
($\rho_s/\rho_{s(\mbox{\tiny vac})}\sim 0$) but still presenting confined quarks ($\Phi=0$). The 
deconfined phase only takes place at $\mu \geqslant \mu_{\mbox{\tiny conf}}$. $\mu_{\mbox{\tiny 
chiral}}$ and $\mu_{\mbox{\tiny conf}}$ are, respectively, the chemical potentials in which the 
broken/restored chiral symmetry and confinement/deconfinement first order phase transitions occur.

The size of the quarkyonic phase in the PNJL0 model is directly governed by the values of the $a_3$ 
and $G_V$ parameters. In Figs.~\ref{phimua3} and~\ref{phimugv1} it is shown that by increasing 
these quantities the quarkyonic phase shrinks. In the case of the~$G_V$ parameter, this situation 
is changed starting from a particular value of~$G_V$. This feature suggested a way to define a 
range of possible $G_V$ values, namely, those producing $\mu_{\mbox{\tiny conf}}=\Lambda$ (the 
cutoff parameter is a energy scale of the model) and $\mu_{\mbox{\tiny 
conf}}=\mu_{\mbox{\tiny chiral}}$. For the case in which $a_3=-0.1$, such a criteria leads to 
$G_V\sim 0.069G_s$ and $G_V\sim 1.18G_s$, respectively, for minimum and maximum values 
of the vector channel strength. 

Finally, we remark the importance of the construction of QCD effective/phenomenological models at 
zero temperature, since a direct application is in the analysis of the hadron-quark phase transition 
in compact neutron stars (described at $T=0$). A recent and very important study evidences the 
existence of quark-matter cores in such objects~\cite{nature}. The challenge of a detailed study 
involving relativistic hadronic models~\cite{rmf,vdw1,vdw2} and the PNJL0 model described here and in 
Ref.~\cite{mattos} is left for a future work.

\section*{ACKNOWLEDGMENTS}
This work is a part of the project INCT-FNA Proc. No. 464898/2014-5, was partially supported by
Conselho Nacional de Desenvolvimento Cient\'ifico e Tecnol\'ogico (CNPq) under grants No. 
310242/2017-7, No. 406958/2018-1 (O.L.) and No. 308486/2015-3 (T.F.), and by Funda\c{c}\~ao de 
Amparo \`a Pesquisa do Estado de S\~ao Paulo (FAPESP) under the thematic projects 2013/26258-4 and 
2017/05660-0 (O.L, T.F.). O.A.M also thanks for fellowships provided by CNPq, INCT-FNA, and
Coordena\c{c}\~ao de Aperfei\c{c}oamento de Pessoal de N\'ivel Superior - Brazil (CAPES - Finance 
Code 001).

\end{document}